\documentstyle[psfig,aas2pp4]{report}

\newcommand{\etal}{{\it et al.\,}}

\slugcomment{To appear in Ap. J.}

\begin{document}

\lefthead{Nichol et al.}
\righthead{Evolution in the X--ray Cluster Luminosity Function Revisited}

\title{Evolution in the X--ray Cluster Luminosity Function Revisited}

\author{R. C. Nichol$^1$ \& B. P. Holden}
\affil{Dept. of Astronomy and Astrophysics, University of Chicago, 5640 S. Ellis Ave, Chicago, Il-60637, USA.} 

\author{A. K. Romer$^1$ \& M. P. Ulmer}
\affil{Dept. of Physics and Astronomy, Northwestern University, 2145 Sheridan Road, Evanston, Il-60208, USA.} 

\author{D. J. Burke}
\affil{Dept. of Physics, University of Durham, South Road, 
Durham, DH1 3LE, UK.}

\author{C. A. Collins}
\affil{Astrophysics Group, School of Electrical Engineering, Electronics and
Phyiscs, Liverpool John Moores University, Bryom Street, Liverpool, L3
3AF, UK.}

$\footnote{Present Address: Department of Physics, Carnegie Mellon University, Pittsburgh, PA-15213, USA.}$

\begin{abstract}

We present new X--ray data taken from the ROSAT PSPC pointing archive
for 21 clusters in the {\it Einstein} Extended Medium Sensitivity
Survey (EMSS).  We have supplemented these data with new optical
follow--up observations found in the literature and overall, 32 of the
original 67 $z>0.14$ EMSS clusters now have new information. Using
this revised sample, we find no systematic difference, as a function
of X--ray flux, between our measured X--ray cluster fluxes and those
in the original EMSS ($\simeq30\%$ scatter).  However, we do detect a
marginal correlation between this observed difference in the flux and
the redshift of the clusters, with the lower redshift systems having a
larger scatter by nearly a factor of two.  We have also determined the
X--ray extent of these re--observed EMSS clusters and find 14 of them
have significant extents compared to the ROSAT Point--Spread Function.
Combining these data with extended clusters seen in the original EMSS
sample, at least 40\% of $z>0.14$ clusters now have an observed x--ray
extent thus justifying their classification as X--ray clusters.

Using our improved EMSS sample, we have re--determined the EMSS X--ray
Cluster Luminosity Function as a function of redshift.  We have
removed potential mis--classifications and included our new
measurements of the clusters X--ray luminosities and redshifts. We
find similar luminosity functions to those originally presented by
Henry \etal (1992); albeit with two important differences. First, we
show that the original low redshift EMSS luminosity function is
insufficiently constrained.  Secondly, the power law shape of our new
determination of the high redshift EMSS luminosity function
($z=0.3\rightarrow0.6$) has a shallower slope than that seen by Henry
\etal. We have compared our new EMSS luminosity functions with those
recently derived from nearby sample of X--ray clusters and find that
the overall degree of observed luminosity function evolution is mild
at best.  This is a result of the shallower slope seen in our EMSS
high redshift luminosity function and a more robust low redshift
determination of the X--ray cluster luminosity function from the
literature.

We have quantified the degree of evolution seen in the X--ray cluster
luminosity using several statistical tests.  The most restrictive
analysis indicates that our low and high redshift EMSS luminosity
functions are statistically different at the $95\%$ level. However,
other tests indicate that these low and high redshift luminosity
functions only differ by as little as $1\sigma$. These data are
therefore, consistent with no evolution in the X--ray cluster
luminosity function out to $z\simeq0.5$.

\keywords{cosmology:observations --  galaxies:clusters:general -- galaxies:evolution -- surveys -- X-rays:galaxies}

\end{abstract}

\section{Introduction}

One of the most striking results in observational cosmology over the
past five years has been the report of rapid evolution in the space
density of X--ray clusters out to $z\sim0.5$.  The strongest evidence
for such evolution has been presented by \cite{gioia90a} \&
\cite{henry92} (H92) using a sample of clusters serendipitously
detected in {\it Einstein} IPC pointings (the {\it Einstein} Extended
Medium Sensitivity Survey: EMSS).  They report that the luminosity
function of X--ray selected clusters steepens significantly as a
function of redshift with the most distant cluster sample having
nearly an order of magnitude less X--ray luminous clusters
($>10^{45}\,{\rm erg\,s^{-1}}$) than the present epoch.

In recent years, observations by the ROSAT X--ray satellite have
focused on extending the redshift baseline of these evolutionary
studies and overall, the results qualitatively agreed with the original
findings of \cite{henry92}. For example, all the specific deep PSPC
pointings towards known optically rich, distant clusters of galaxies
($z>0.6$) have yet to discover a single X--ray bright cluster
(\cite{nichol94}, \cite{castander94} \& \cite{bower94}).  These
studies however, may be biased since most of their original targets
were drawn from optically selected catalogues.

In the long term, such bias can be removed by compiling serendipitous
samples of distant X--ray clusters from the large archive of ROSAT
PSPC pointings in the same vein as the EMSS.  Several groups have
already begun this task and their combined effort should provide a
powerful database for quantifying the degree of observed X--ray
cluster evolution (\cite{castander95}, \cite{rosati95},
\cite{burke95}, \cite{scharf97} \& \cite{romer96}). This work is still in its infancy,
mainly because of the large amount of optical follow--up needed to
verify the presence of a distant X--ray cluster and to exclude
possible contaminants like stars and AGNs.  Preliminary
results from these ROSAT surveys either appear to be consistent with the 
EMSS sample (see \cite{castander95}), or, advocating no evolution
(see \cite{collins97}). 

The theoretical consequences of X--ray cluster luminosity evolution
are significant, suggesting that the large--scale structure in the
universe evolved hierarchically over a relatively short look--back
time.  However, the present results are inconsistent with simple
scale--invariant hierarchical clustering models, since these actually
predict an increase in X--ray bright clusters with increasing redshift
(\cite{kaiser86}).  Several authors have resolved this discrepancy but
at the expense of breaking scale--invariance and allowing significant
energy input into the gas at high redshift (\cite{kaiser91},
\cite{Evrard91}, \cite{cavaliere93}). Alternatively, the problem could
be alleviated by steepening the slope of the initial density power
spectrum on cluster scales, yet this would be inconsistent with
popular Cold Dark Matter theories of structure formation
(\cite{davis85}, \cite{henry92}).

Considering the underlying
theoretical importance of cluster evolution, we report
here a re--examination of the EMSS cluster sample since this database
is the most robust, well--understood catalogue of X--ray clusters
out to $z\simeq0.5$ presently available.  The EMSS, as a whole, has received
intense optical scrutiny over the past decade and many of its 835
X--ray sources now have a secure identification (\cite{gioia90b},
\cite{stocke91}). In the next section, we describe new data available
from the ROSAT pointing archive on nearly half of the $z>0.14$ EMSS
clusters used by H92 in their evolutionary study.  These new data therefore,
provide an important check since many of these distant EMSS clusters
were only $\simeq5$ sigma detections in the {\it Einstein} data, {\it
i.e.} there is a large uncertainty on their X--ray fluxes.  The ROSAT
re--observations of these clusters represent a significant improvement
on this for three reasons: First, the clusters received, on average,
longer exposure times; second, ROSAT has a much lower particle
background thus improving its sensitivity to low surface brightness,
extended objects; third, the satellite has greater on--axis angular
resolution.

This new ROSAT dataset has allowed us to test two critical concerns
about the EMSS cluster sample: {\it i)} the accuracy of the X--ray
cluster fluxes and therefore, their luminosities; {\it ii)} the
classification of the X--ray emission as a hot, intracluster medium.
The latter point is very important with regards to identication since
X--ray point sources such as stars and AGNs dominate the X--ray source
counts.  Therefore, only a small error in classifying these objects as
clusters, and vice versa, will make a large difference to the final
X--ray cluster subsample. Therefore, the most secure signature of
cluster emission is X--ray extent.  This re--examination of the EMSS
is part of a large project aimed at searching the ROSAT data archives
for serendipitous observations of distant X--ray clusters;
Serendipitous High--redshift Archival ROSAT Cluster (SHARC) survey
(see \cite{burke95}, \cite{ulmer95}).  Throughout this paper we assume
$H_o=50\,{\rm km\,s^{-1}\,Mpc^{-1}}$ and $q_o=0.5$ to be consistent
with previous work.

\section{New Data and Analysis}

We present, in Table 1, all the $z>0.14$ EMSS clusters, from Table 2
of H92, for which new data are available including re-observations by the ROSAT
satellite and/or further optical identification work by Gioia \& Luppino (1994; GL94) and others.  Column
1 gives the EMSS name of the clusters followed by, in column 2, the
ROSAT pointing identification number.  Column 3 presents the measured
redshift of the clusters which were taken from GL94 or Carlberg \etal (1996). Columns
4, 5 \& 6 give the results of our X--ray source analysis and indicates
whether we find the cluster X--ray emission extended or not (the numbering given in column 6 is the same as that in Figures 1 and 2). Column 7
gives the PSPC off--axis angle (in pixels) of the clusters and
indicates cases where the cluster was the original requested target of
ROSAT; $\simeq85\%$ of our ROSAT EMSS cluster re--observations were
again accidental. Column 8 is the averaged PSPC exposure time (vignetted
corrected), while column 9 gives the net source counts (within the
bandpass $0.4\rightarrow2.0\,$keV).  Many of the re--observed EMSS
clusters are detected to significantly higher signal--to--noises in
our data than that used by H92. Columns 10 \& 11 are the measured
source and background X--ray count rates and the final four columns
are the X--ray fluxes and total luminosities computed from these
observed count rates in both the ROSAT and {\it Einstein} bandpasses.

The PSPC pointing data used in this paper were obtained from the GSFC
ROSAT data archive and reduced using a suite of programs we have
developed as part of the SHARC survey. We refer the reader to our
forthcoming paper,
\cite{romer96}, that describes in detail our data reduction
techniques. However, for completeness, we briefly highlight the main
features of our analysis here.  The raw X--ray data were initially
reduced using the methodology outlined by \cite{snowden94}. For each
pointing, an individual exposure map was constructed from the
satellite's orbital information and divided into the binned X--ray
photon data (15 arcsecond pixels).  We concentrated on the ``hard''
ROSAT band, $0.4\rightarrow 2.0\,$keV, since this substantially reduces
the background count rate. The exposure and vignetted corrected images
were then analysed using our own source detection algorithm which is
based on a mexican--hat wavelet transform.  This particular wavelet is
well--suited to analysing PSPC data since it is the second derivative
of a Gaussian and therefore, this underlying function matches closely the
(radially averaged) shape of the ROSAT Point Spread Function (PSF)
which can be approximated by a Gaussian.

A crucial part of our re--analysis of the EMSS sample was to determine
if the observed X--ray cluster emission was extended or not. For this
reason, we present this particular stage of the analysis in greater
detail.  The extent of all sources in a pointing was determined
directly from our wavelet data taking advantage of the fact that 
the width of the mexican hat filter, at it's zero--crossing points, 
is equal to the width (between inflection points) 
of the original Gaussian it was constructed from.
Therefore, we grew each source outwards from its initial centroid
until we reached zero in wavelet--space. This source boundary, in
wavelet--space, is then a measurement of the size of that source and
is simply the product of the size of the mexican--hat (we
define this as $\sigma_w$ which is the dispersion of the underlying
Gaussian) and the size of the observed source, which in most cases
will be the PSF.  A final centroid analysis is then carried out on
each candidate source, using only the pixels within this
zero--crossing boundary, and their major and minor axes determined.

This procedure was independently carried out for three different
mexican--hat sizes; $\sigma_w=$ 3, 6 and 9 pixels.  The power of this
approach is that the different size filters are sensitive to different
regimes of the off-axis PSF. The smallest filter is best suited to the
inner region of the PSPC field--of--view, while further off--axis, the
larger filter sizes become more powerful as the PSF increases. For
example, at large off-axis angles, the smallest filter tends to
break--up sources because the surface brightness profiles of sources
flattens due to increases in the PSF and reduction in the
ROSAT telescope effective collecting area. The $\sigma_w=$ 6 and 9 pixel
filters help to recover from this.  Therefore, by changing the filter
size in our analysis, we naturally accommodate for a changing PSF thus
allowing us to probe further out into the ROSAT field--of--view.

Figures 1 and 2 shows the distribution of the major and minor axes, as
a function of off-axis angle, for 1637 high signal--to--noise 
sources detected in the
$\sigma_w=3$ wavelet maps of 407 deep ROSAT PSPC pointings.  
The size of the these sources is the convolution of the wavelet size
with the intrinsic source size and therefore, they are larger
expected.
These pointings
have been reduced as part of the SHARC survey which is presently
investigating all extended sources seen in these data.  The other
$\sigma_w$ maps are very similar and are therefore, not shown.  The
only difference, as highlighted above, is the area over which the PSF
locus is well--defined has been shifted to higher off-axis angles.
Clearly visible in Figure 1 is the expected increase in the ROSAT PSF
with off--axis angle above which we must determine if a source is
extended or not.  Also shown is a line which represents sources that
are $>3\sigma$ extended and was empricially determined, as a function
of off--axis angle, from these sources (see \cite{romer96} for
further information).

The off--axis angle and major and minor axes of all EMSS clusters
detected in the 3 pixel wavelet map are presented in Table 1 (column 6) 
and are
plotted in Figures 1 and 2. An EMSS cluster is designated as extended if
its major and/or minor axis is above this $3\sigma$ line in any of the
3 wavelet maps, for example, MS0811.6+6301 was selected as extended
based solely on the $\sigma_w=6$ wavelet information. A question mark
in Table 1 signifies clusters that do not satisfy these criteria, but
appear to lie away from the point--source locus in two or
more of its wavelet representations.

The X--ray count rates of the EMSS clusters were measured
interactively using the IRAF/PROS software package. This differs from
the approach used in the SHARC survey and therefore, we describe it
fully here.  In the original EMSS, the cluster count rates were
computed within a fixed angular aperture ($2.4\times2.4$ arcmins) and
were later corrected, via a cluster model, to the total X--ray flux.
We, however, had the freedom to select a fixed metric aperture thus scaling 
it sppropriately with redshift and the PSF for each individual cluster
(we convolved the angular size of the metric aperture with the expected size
of the PSF). For aa assumed King profile with a core radius of
$0.25{\rm Mpc}$ and $\beta=\frac{2}{3}$, we computed the angular size
of an aperture for each cluster that would encompass 85\% of the total
cluster flux, irrespective of redshift and off--axis angle (the fluxes
and luminosities given in Table 1 are corrected for the 15\% of light expected
outside the aperture).
Furthermore, the interactive approach allowed us examine the
environment of each cluster and exclude any nearby sources (both in
the background and source apertures).  Wherever possible, the
background was estimated from an annulus around the source
aperture having a similar number of pixels as the source
aperture.  If this was not possible, a nearby area devoid of sources
was used, ensuring that it was at the same off-axis distance as the
source so the vignetting corrections would be comparable.

These measured count rates were finally converted to fluxes (${\rm
erg\,cm^{-2}\,s^{-1}}$) by integrating a 6keV Raymond--Smith (RS)
spectrum through the ROSAT response function and accounting for
absorption using HI column densities interpolated from Stark {\it et
al.} (1992) and the model by Morrison \& McCammon (1983). This
therefore, gave us the X--ray cluster flux in the ROSAT bandpass
outside our own Galaxy.  We then converted the ROSAT fluxes to that
expected in the {\it Einstein} bandpass again using a 6keV RS
spectrum. We chose this model cluster spectrum to be fully consistent
with the H92 analysis. The cluster
luminosities (including 
k--corrections) for both the ROSAT and EMSS
bandpasses are given in columns 14 \& 15 of Table 1 respectively.

Ideally, we would like to convert our ROSAT count rates to {\it
Einstein} fluxes using the individual spectra of these EMSS clusters.
However, for a vast majority of the EMSS such data does not exist.
Therefore, to quantify this potential systematic effect, we
re--computed the expected {\it Einstein} fluxes of all our ROSAT EMSS
detections using the Luminosity--Temperature relationship of
\cite{wang93} and iterated it until we gained a stable solution
(within 0.1\% between iterations).  This usually took $<4$ passes and
always changed the final iterated cluster luminosity by less than 6\%.
For example, MS0015.9+1609 shifted from ${\rm L_x(44)}=22.78\,{\rm
erg\,s^{-1}}$ -- which is given in Table 1 -- to ${\rm
L_x(44)}=23.72\,{\rm erg\,s^{-1}}$ for an iterated temperature of
11.35keV. Our count rate to flux conversion was also relatively
insensitive to variations in the RS metal abundance and varied by less
than 10\% over a reasonable range of metallicities.

In Figures 3 and 4, we present all available ROSAT data on the distant
$z>0.14$ EMSS clusters discussed in this paper. The size of the
individual plotting windows in Figure 3 is equal to the diameter of
the aperture used to compute the cluster count rates.  Also shown is
the measured position angle of the observed x--ray emission and the
radial direction towards the center of the PSPC.  In Figure 4, we
present the ROSAT radial X--ray surface brightness profiles for the
clusters as well as the appropriate off-axis ROSAT PSF. For the PSPC
data, this was computed from the point--source locii seen in Figures 1
and 2 (not the $3\sigma$ line). For the HRI data, we used the model of
David \etal (1993) for the PSF. These plots clearly show that many of
the re--observed EMSS clusters are high signal--to--noise X--ray
detections and are extended.

\addtocounter{table}{1}
\begin{table*}[tbph]
\begin{center}
\caption{
Maximum likelihood Parameterisation of XCLFs derived from
the original EMSS Data Sample}
\begin{tabular}[]{c|ccc} \hline
Redshift shell & Number & $\alpha$ & $K (10^7 Mpc^{-3}L_{44}^{\alpha-1})$ \\ \hline 
0.14 - 0.2& 21 & $2.24\pm0.32$ ($2.19\pm0.21$) & $7.5\pm1.9$  ($5.85\pm0.25$) \\ 
0.2 - 0.3 & 23 & $2.72\pm0.29$ ($2.67\pm0.26$) & $7.6\pm1.9$  ($6.82\pm0.51$) \\ 
0.3 - 0.6 & 23 & $3.37\pm0.31$ ($3.27\pm0.29$) & $14.8\pm5.2$ ($12.33\pm3.87$) \\ \hline 
\end{tabular}
\end{center}
\end{table*}

\section{Comparison of Cluster X--ray Flux Measurements}

One of the motivations of our work was to check the robustness of the
EMSS flux measurements. In H92, the total flux was computed, using a
King model, from the detected flux within a fixed angular aperture.
In contrast, we attempt to measure the total flux directly using a fixed 
metric
aperture.  For the
21 ROSAT--EMSS clusters, Figure 5 presents the difference  
between our ROSAT mesurement of the total flux of the EMSS clusters 
in the {\it Einstein} bandpass (column 13 in Table 1) and 
that given by H92, versus our total
ROSAT--EMSS flux. 
The error bars plotted in Figure 5 were derived from the propagation of
the errors published on the net counts by H92 and in Table 1 of this
paper (column 9). Using a $\chi^2$ test, we have compared the data in Figure 5
with the expected relationship {\it i.e.} the zero line in Figure 5. 
For all the data points, we find $\chi^2=57$ for 20 degrees of
freedom (d.o.f) which is highly improbable. However, much of this
disagreement is due to the two points at the highest total flux
(MS0451.5+0250 and MS0735.6+7421).  Therefore, if we constrain our
analysis to ${\rm F_x<2\times10^{-12}\,erg\,s^{-1}\,cm^2}$ and ${\rm
F_x<10^{-12}\,erg\,s^{-1}\,cm^2}$, we find $\chi^2=33$ (18 d.o.f) and
$\chi^2=12$ (12 d.o.f) respectively. These indicate that the observed 
scatter (28\%) at lower flux levels is consistent with the observed 
errors on the two datasets (ours and H92).

\begin{figure}[tbhp]
\centerline{\psfig{file=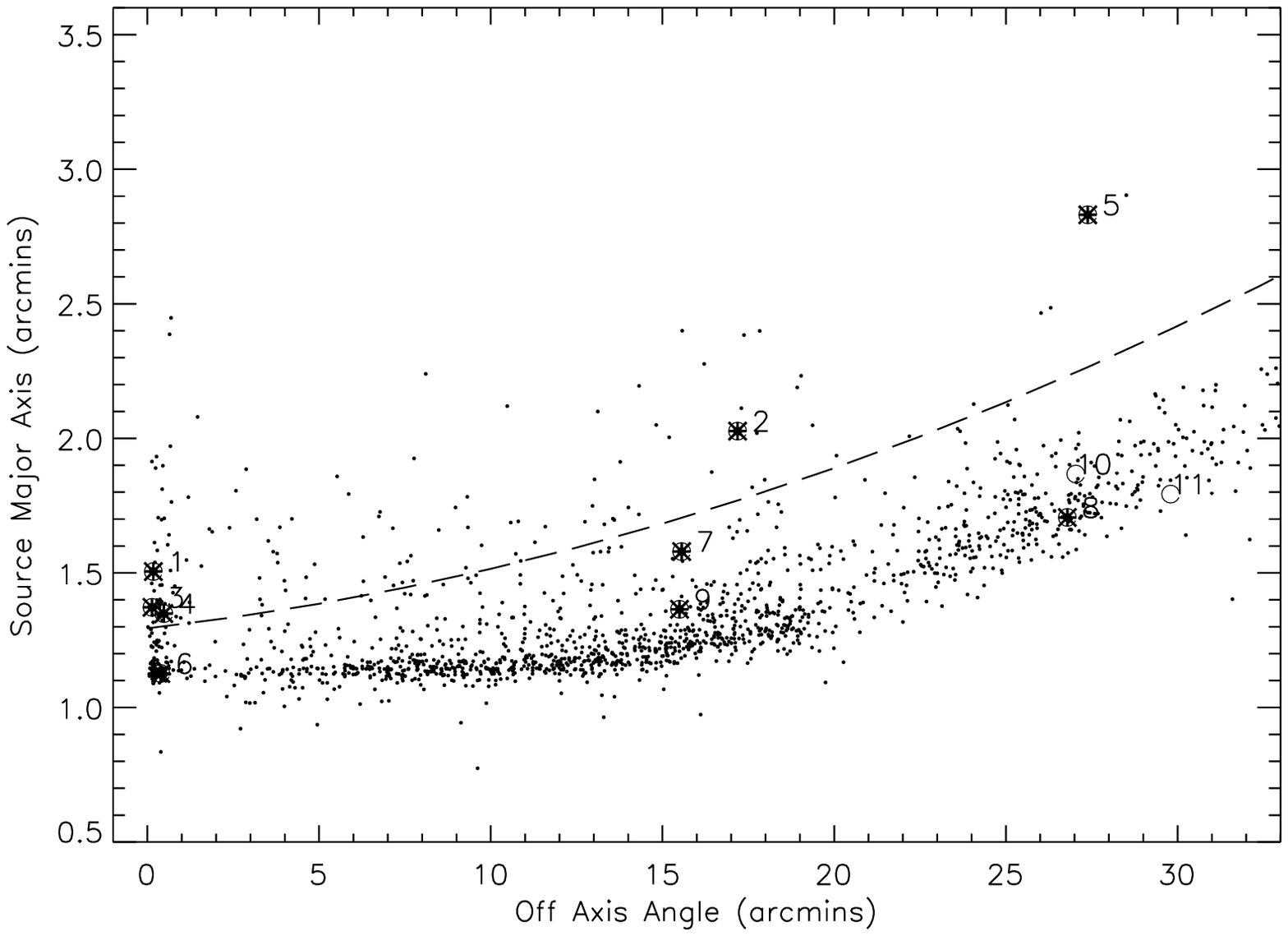,height=3in}}
\centerline{\psfig{file=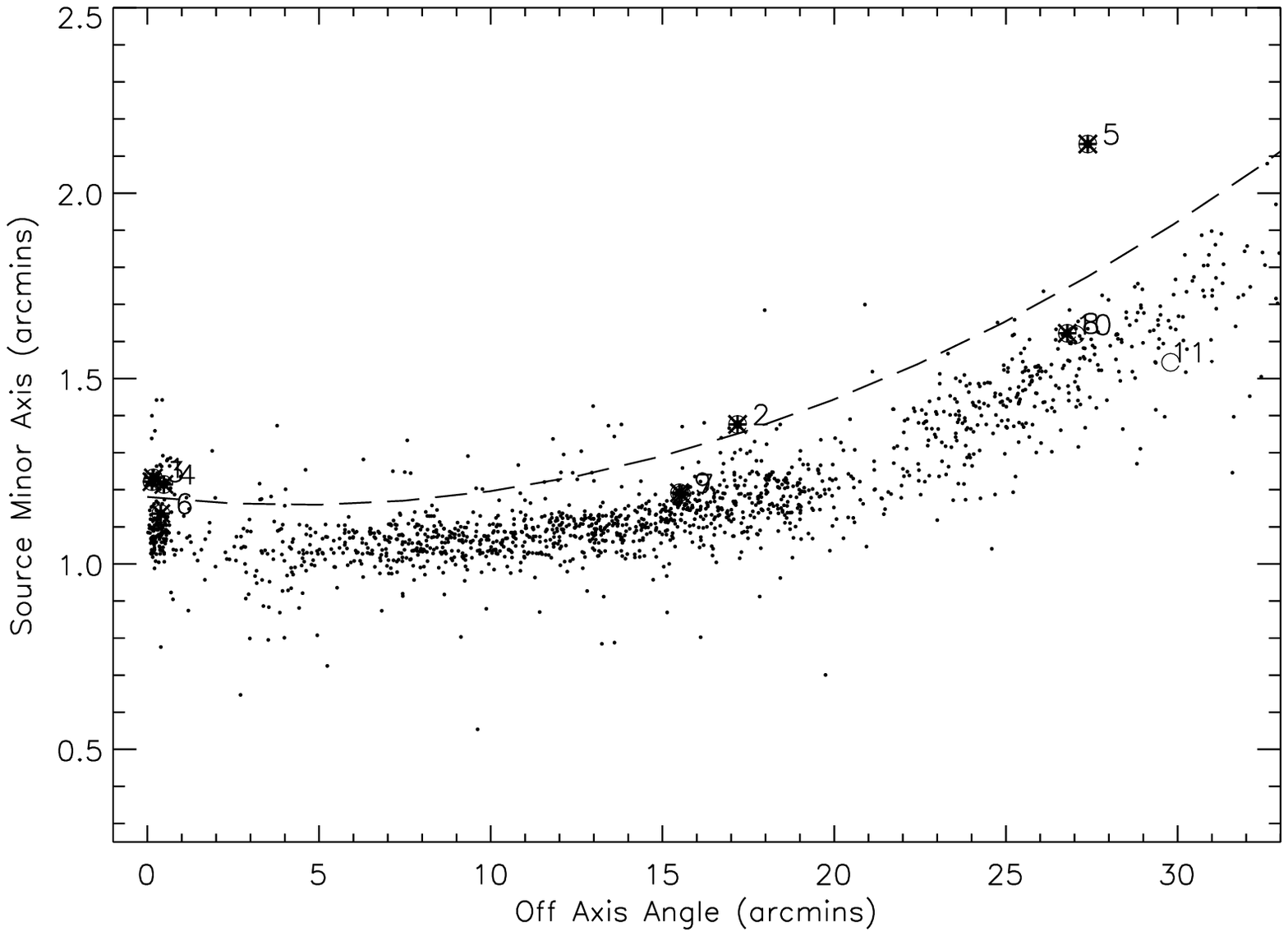,height=3in}}
\caption{The major and minor axes for 1637 high signal--to--noise 
sources detected in 407
deep ROSAT PSPC pointings versus their off-axis angle (the x--axis
the convolution of the source size with the wavelet filter size). 
The change in the PSPC PSF with off-axis angle is visible and was defined
empiricially (the dotted lines represent $>3\sigma$ extended sources).
The star symbols are $z>0.3$ EMSS clusters and have the same numbering
as given in column 4 of Table 1.  The unfilled circles are the 2 EMSS clusters
that only have lower limits measured for their extents. These sources
were found in the $3$ pixel wavelet map.} 
\end{figure}

\begin{figure}[tbhp]
\centerline{\psfig{file=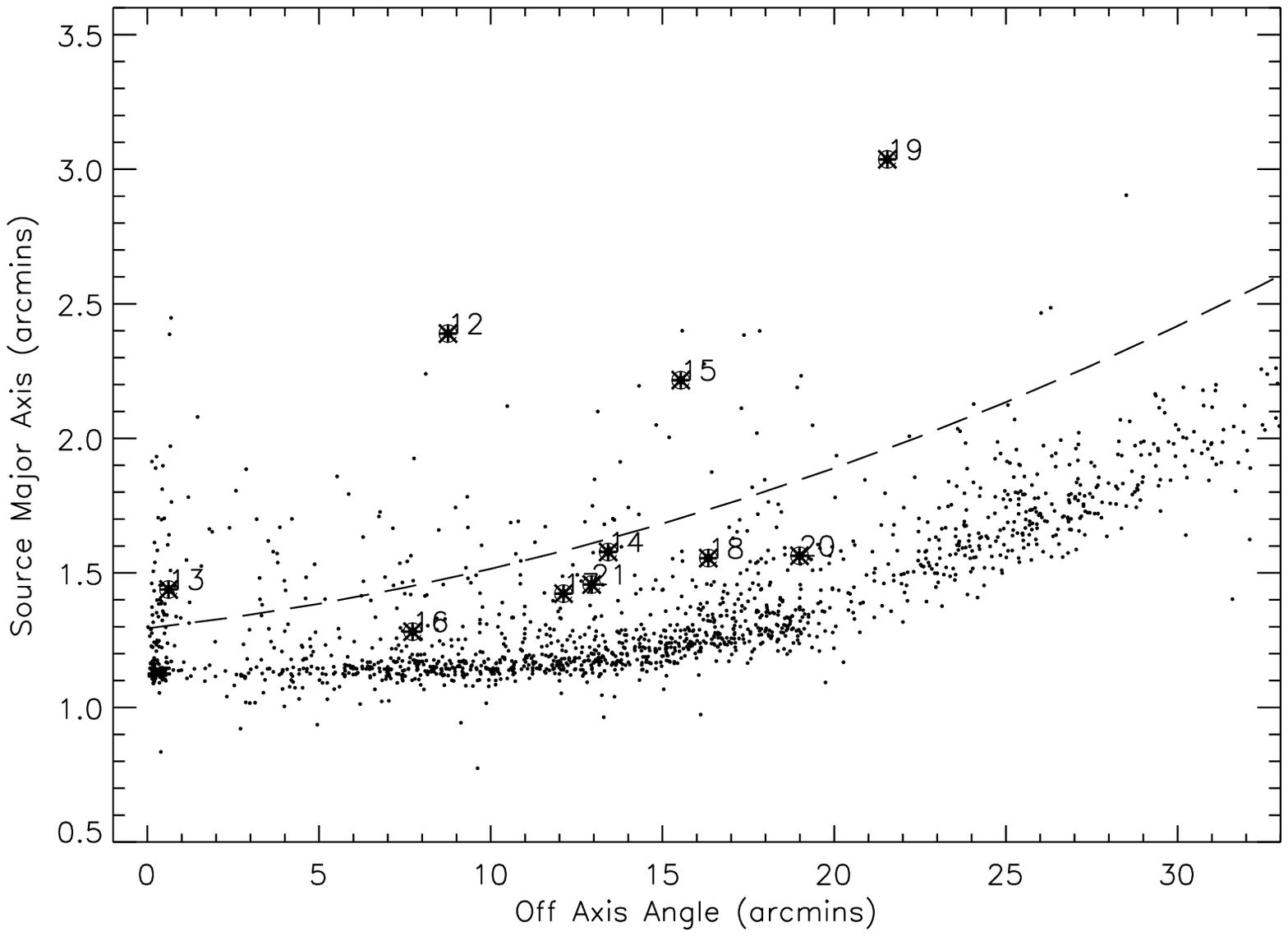,height=3.in}}
\centerline{\psfig{file=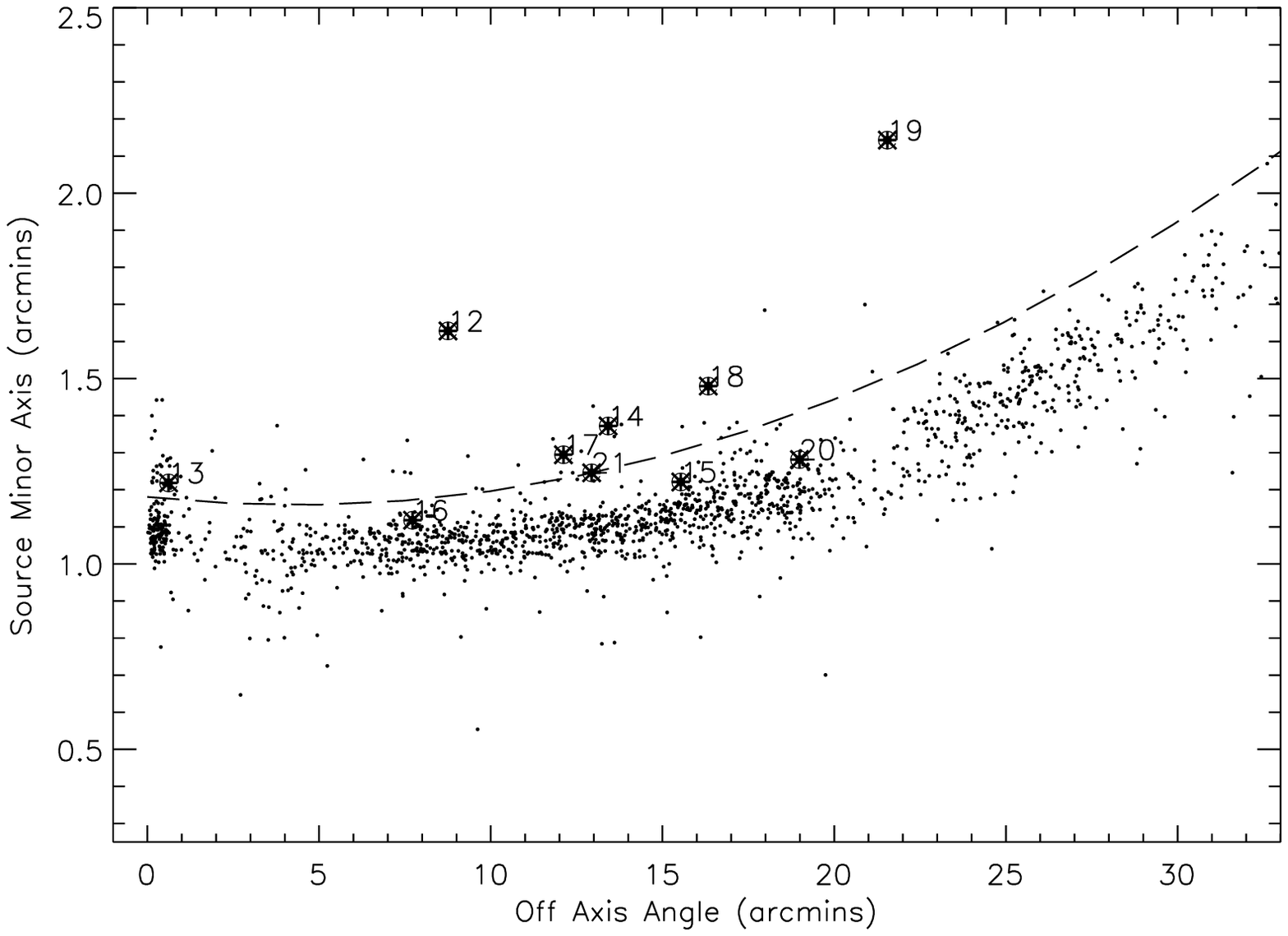,height=3.in}}
\caption{The same as Figure 1 but showing the $z<0.3$ EMSS
clusters (solid symbols). Again, the numbers coincide with those given
in column 4 of Table 1.} 
\end{figure}

A more interesting plot to make is to plot the flux differences
in Figure 5 as a function of redshift. This is shown in Figure 6.  In
this case, there does appear to be a noticeable decrease in the flux
differences with increased redshift.  This is quantified by splitting
the sample at $z=0.3$, with the lower redshift clusters having a mean
absolute difference of 39\%, while for the higher redshift data, the
mean absolute difference is only 19\%. The standard error on both
these means is $\sim30\%$, since both were determined using $\simeq10$
clusters. Therefore, this observed drop in the mean absolute flux
difference is marginally significant.  Such a decrease with redshift
is consistent with the H92 methodology since at higher redshifts there
would be a greater amount of the clusters' total flux within the fixed
angular aperture compared to lower redshifts, thus the required
correction would be smaller.

\section{Measurement of Extended Cluster X--ray Emission}

It is imperative that the classification of all EMSS clusters should
be made as secure as possible. This is straightforward using the
measured extent of the observed X--ray emission from the cluster.
Figures 1 and 2 present the observed extent of our ROSAT--EMSS
clusters against that expected for X--ray point--sources (as a
function of off-axis angle).  Overall, 14 of the 21 clusters with new
ROSAT data are flagged extended in Table 1 (column
4). For 3 of these extended sources, H92 also detected an extent in
the {\it Einstein} data (these clusters are highlighted in Table 1).
By combining our extent results with H92, a total of 25 $z>0.14$ EMSS
clusters now have sufficient X--ray information to measure a significant
X--ray extent ($40\%$ of the whole EMSS sample).  Constraining
ourselves to $z>0.3$, this figure drops to $30\%$. For these extended
clusters, therefore, we are most probably seeing X--ray emission from
a hot, intracluster gas which undoubably strengthens their
classification as X--ray clusters.

Next we discuss the clusters marked as
non--extended, or uncertain, in our database. Some of these are the
result of low signal--to--noise, low intrinsic X--ray luminosity
and/or uncertainties in the PSF at large off--axis angles.  For
example, MS0418.3--3844 is very
close to the ROSAT rib structure resulting in an unrealistic
major and minor axis in all 3 wavelet maps and therefore, we only
present a lower limit to the cluster size in Table 1.  The same
problem occurred for MS811.6+6301 in the $\sigma_w=3$ wavelet map, but
it was observed to be extended in the larger wavelet maps.

For EMSS clusters MS1209.0+3917, MS1219.9+7542, MS1512.4+3647 and
MS2137.3-2353, there exists significant archival ROSAT HRI data which
can help classify these systems. In the PSPC data, they all lie on --
or near -- the observed point source locus in all three wavelet maps
(see Figures 1 and 2). They also appear unextended in Figure 4  
Considering the length of exposure time
gained in these PSPC data, we suggest that these clusters either have
compact cores, or, that they are mis-identifications.  For MS1209.0+3917
and MS1219.9+7542, the HRI indicates that these X--ray sources are
point--like with little evidence for any X--ray extent.  The optical CCD image
of MS1209.0+3917 by GL94, and their own suspicions, supports the idea that
this cluster is a mis-identification. For MS1219.9+7542, this system is a 
poor group with a dominant galaxy showing moderate [OII] emission in its' 
optical spectrum (GL94 and Stocke
\etal 1991). 
The X--ray and optical data combined, appears to 
indicate that the x-ray emission for these two EMSS clusters may be due to 
an AGN and not a hot, intracluster medium. We therefore, remove these 
clusters in any further analysis of the EMSS sample.

For MS2137.3-2353 and MS1512.5+3647, the HRI data indicates that the X--ray
emission from these systems is extended compared to the PSF.  We
fitted a King model to the HRI data and measured core radii of
$17\pm8$ ($\beta=0.77\pm0.28$) and $16\pm10$ ($\beta=0.63\pm0.23$)
arcseconds for MS2137.3-2353 and MS1512.5+3647 respectively.  Clearly,
both clusters could not have been expected to be resolved by the PSPC
instrument since even on-axis, the PSPC PSF has a $\simeq20$
arcseconds FWHM (illustrated in Figure 4).  For both systems, the
angular extent of the X--ray emission represents a metric core
radius of less than $100$ kpc; significantly smaller than 
the rest of the EMSS sample.  For MS2137.3-2353, the lensing work of Mellier
\etal (1993) agrees with this metric core radius, while the optical appearance
of this system is dominated by a single, large cD--like galaxy with a 
similar metric size as the observed X--ray emission (GL94).  For MS1512.5+3647, Stocke
\etal (1991) has classified this system as a ``cooling--flow galaxy''
which is defined to be a galaxy with similar optical characteristics
as the dominant galaxy in a cooling flow clusters but which is no
associated with a rich, optical cluster of galaxies (see the CCD image in 
GL94).
Carlberg \etal 1996 has extensive redshift information on this EMSS
cluster and measure a velocity dispersion of $\sigma_v=690\,{\rm
km \, s^{-1}}$. However, this system appears to have substantial redshift
contamination (as seen in Figure 1 of their paper) and may indicate that
MS1512.5+3647 is a projection effect.  More data are required on these
two systems to unambiguously classify them.  We do not remove
MS2137.3-2353 and MS1512.5+3647 in our later analysis, but note that our
results were quite insensitive to their inclusion, or, removal.

\begin{figure*}[thp]
\centerline{\psfig{file=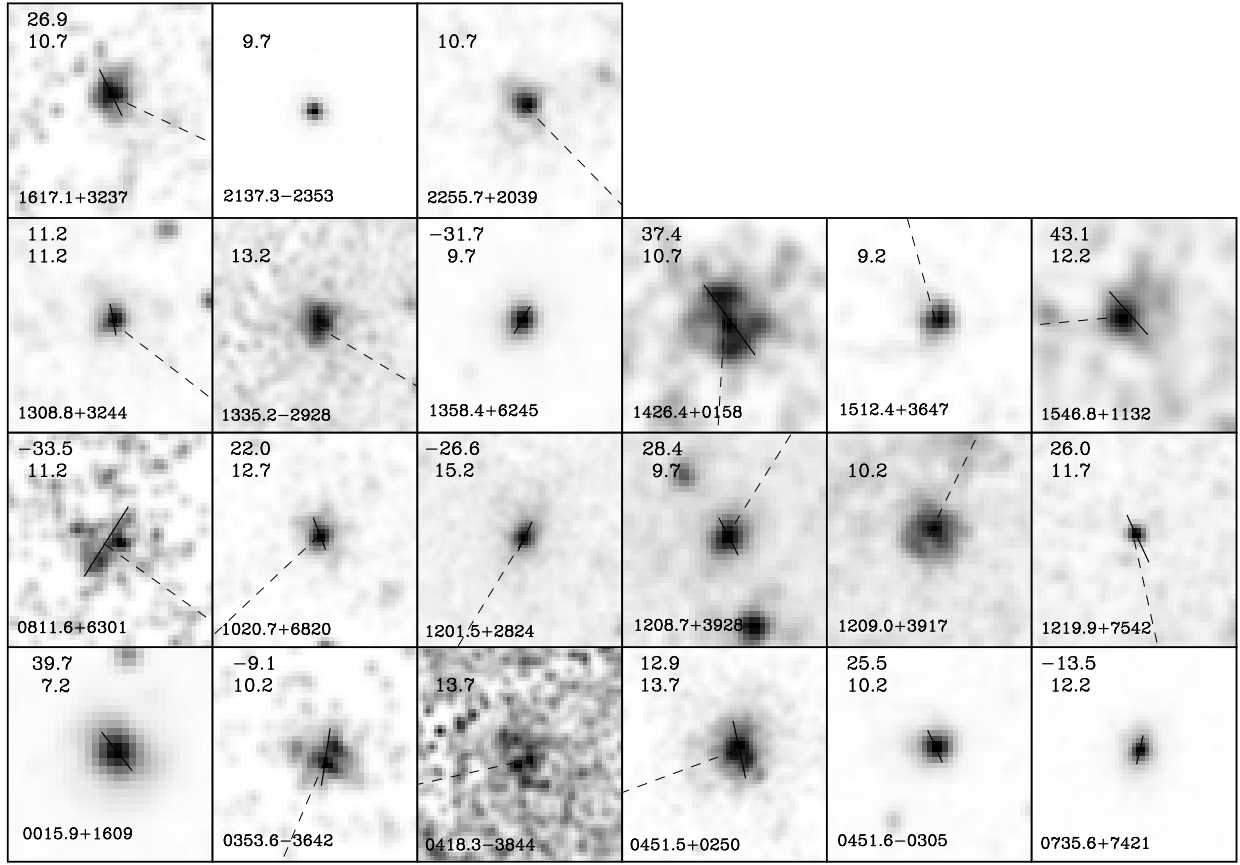,height=5.0in,angle=270}}
\caption{A greyscale plot of the count rates of the 21 EMSS cluster 
with new ROSAT data given in Table 1. North is upwards with East to the left.
The solid line represents the measured position angle of the cluster
and is only computed for clusters with ${\rm minor/major}>0.8$, thus
avoiding circular clusters. The numerical value -- in degrees -- of
the position angle is the top number shown in the top--left hand
corner of the window. The dashed line is the direction towards the
center of the PSPC and helps the reader determine if the asymmetrical
PSPC PSF has influenced the determination of the position angle. In
some cases, this looks likely. The dashed line is not plotted if the cluster
was nearly on-axis.  The count rate data are scaled linearly between
zero and the maximum observed count rate (these maps have been
vignetted and exposure corrected). The contribution of the PSF has not
been removed and these images have also been lightly smoothed with a
Gaussian of width one pixel to reduce the binning noise.  The size of
the plotting window is given as the bottom number in the top
left--hand corner and is equal to the aperture used in the flux
determination. Other sources near the cluster were removed prior to
any flux determination {\it i.e.} MS1208.7+3928.}
\end{figure*}

No high quality ROSAT HRI data exists for the remaining 2 clusters 
marked as non--extended, or uncertain, in Table 1 
(MS1208.7+3928\footnote{
MS1208.7+3928 was not detected in the ROSAT HRI pointing
701844 towards MS1209.0+3917.} \& MS1617.1+3237). For these two, we therefore
performed Monte Carlo simulations to assess whether we had the
required signal--to--noise to observe an extent or not (these
simulations are presently being developed as part of the SHARC survey
to help characterise the selection function of that survey and will be
presented in detail in a forthcoming paper). Briefly, the simulations
used here involved adding false clusters to the relevent ROSAT PSPC 
pointings (Table 1)
using an underlying King profile of core radius $0.25$ Mpc and
$\beta=\frac{2}{3}$. These false clusters were given the same
signal--to--noise as the real real detections and were positioned 
at the same off--axis angle (usually opposite the real cluster in the PSPC
field--of--view). For each cluster, 10 iterations were
constructed and passed through our source detection software.
This allowed us to
determine the frequency with which these false sources were flagged as
extented.
Using these simulations as a guideline, it would appear that
MS1208.7+3928 is more compact than expected since it was always 
detected as extended in the simulations. MS1617.1+3237, however, remains
uncertain since the simulations indicated that we do not have high enough
signal--to--noise to unambigously detect an extent given this nominal
profile (it was flagged as extended only $\simeq 70\%$ of the time).

\section{The X--ray Cluster Luminosity Function}

\subsection{A Re--Determination of the H92 Result}

As mentioned earlier, the most striking result to come out of the EMSS
cluster sample is evidence for rapid negative evolution in the X--ray
Cluster Luminosity Function (XCLF; H92). Considering the amount of new
information gained on the EMSS sample of clusters over the past few
years, we have revisited the H92 result here. Our motivation for this
are two--fold: First, to determine if actual changes in specific
clusters make a significant systematic difference to the observed
XCLF; secondly, to assess the robustness of the result to small, yet
significant, changes in the overall cluster population.  For instance,
if the XCLF was radically altered, in any sense, by changes to a few
clusters, it would indicate that the sample was not stable and remove
much of its predictive power. In conjunction, we have re-assessed the
statistical significance of X--ray evolution originally reported by H92.
 
\begin{figure*}[thpb]
\centerline{\psfig{file=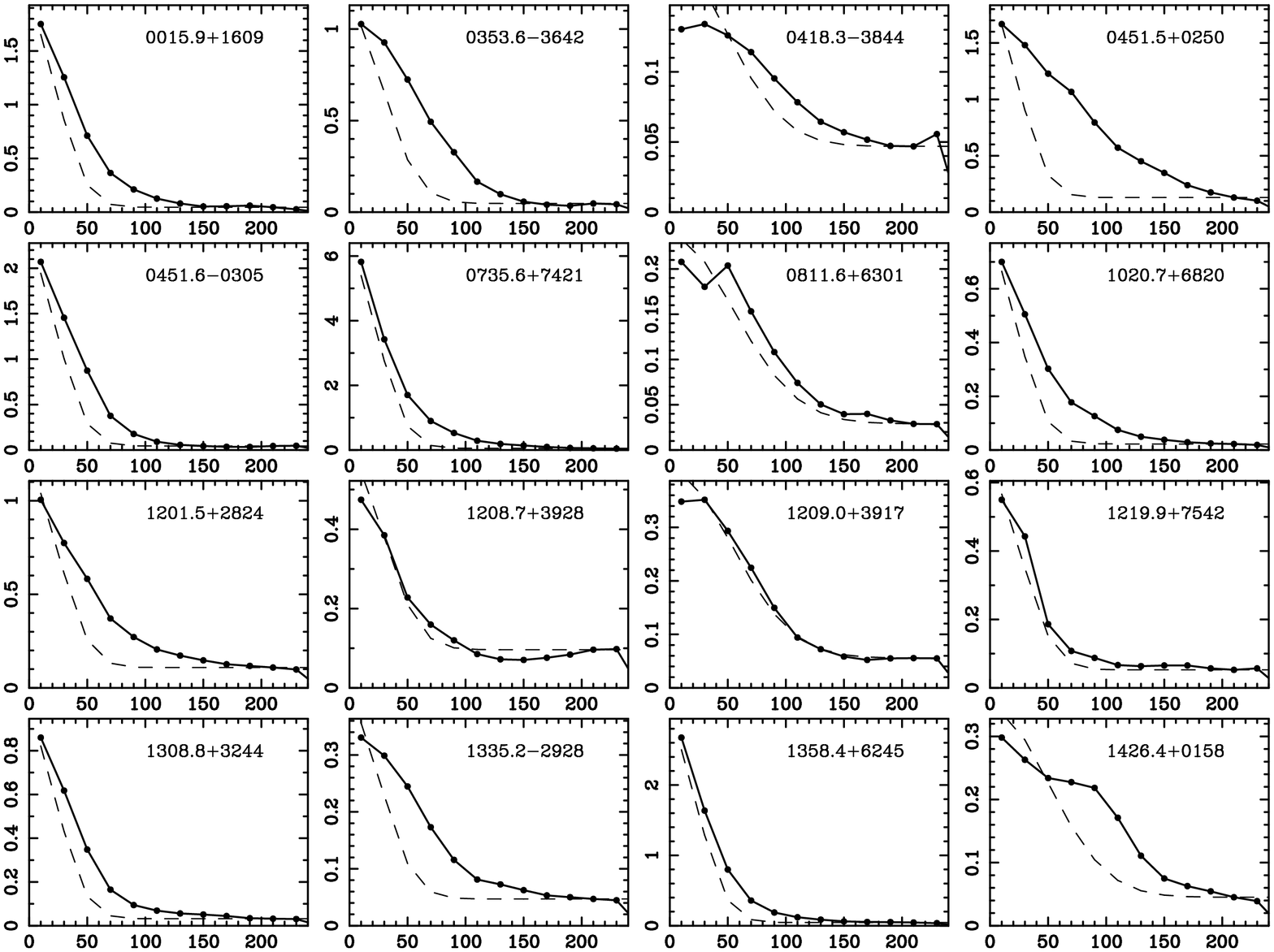,height=3.5in,angle=270}}
\centerline{\psfig{file=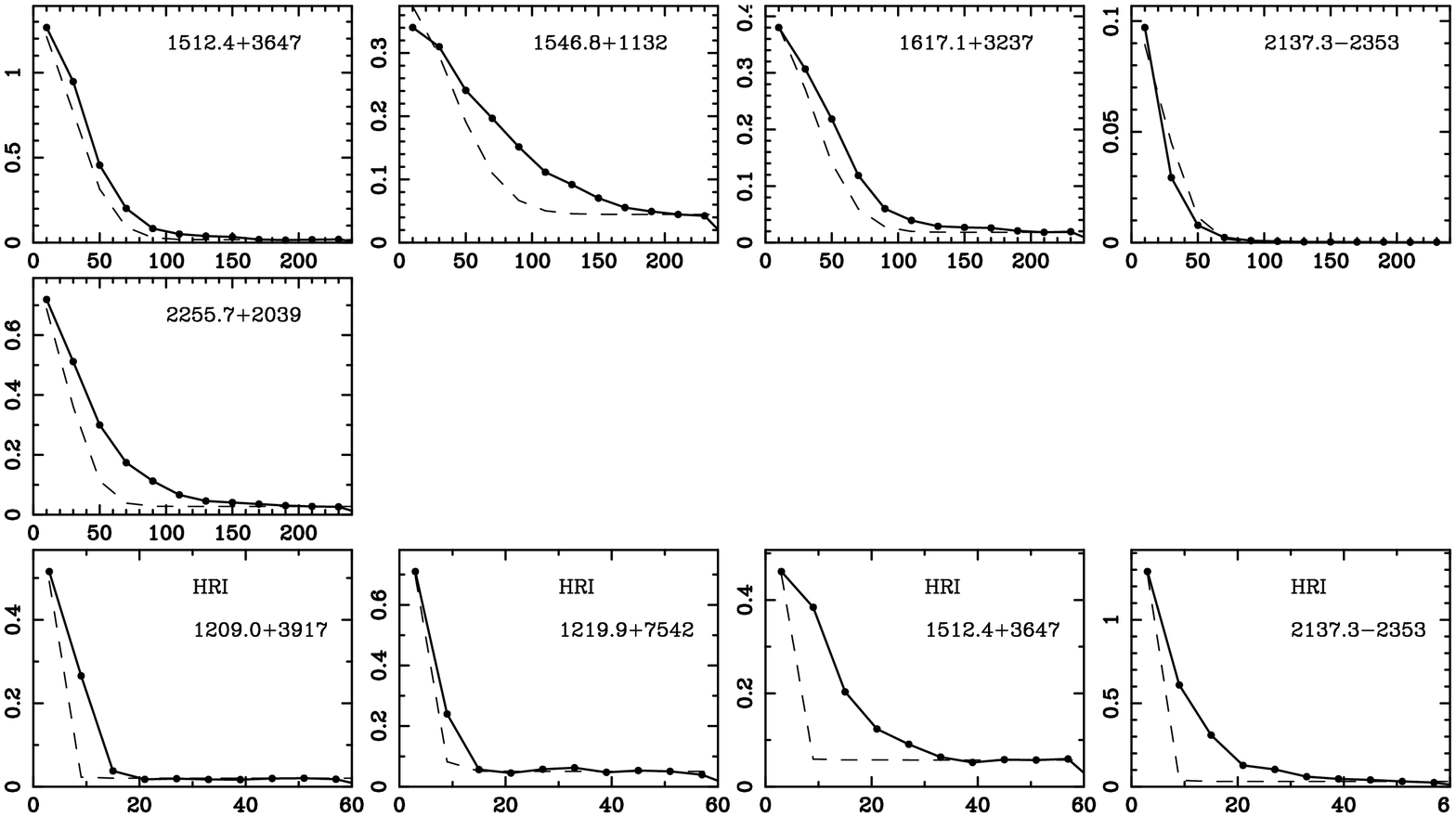,height=3.5in,angle=270}}
\caption{
The radially--averaged
profiles for 21 EMSS clusters with new ROSAT PSPC data (given in Table 1) and 4 EMSS clusters with available HRI data.
The x--axis is in arcseconds from the intensity--weighted centroids
of the clusters, while the y--axis is the observed X--ray surface brightness
(${\rm counts\, s^{-1}\, arcsec^{-2}}$). The binsize used for the PSPC and HRI 
data was 20 and 6 arcseconds respectively. The dashed lines show the 
expected size of the PSF at that given off-axis angle (as given in Table 1).
The PSPC PSF was empirically determined from the data given in Figures 1 and 2
(we use the observed point--source locus not the $>3\sigma$ line shown). 
The HRI PSF was taken from David \etal 1993}
\end{figure*}

We initially re--computed the XCLF using the same $1/V_{a}$
test as employed by H92 (\cite{avni80}).  This involved
computing the maximum observable volume ($V_a$) for each cluster,

\begin{equation}
V_{a} = \sum_{i=1}^{N} \Omega_{i}\Delta V(z_{max},z_{min}),
\end{equation} 

\noindent where $\Omega_i$ is the area of the EMSS survey to a given 
sensitivity limit (see Table 3 in H92), $N$ is the total number of different sensitivity limits (19 in total from Table 3 in H92)  and $z_{min}$ is the lower redshift
limit for a given redshift shell. For
\( z_{max} \), the smaller of either the maximum detectable 
redshift for a given cluster, or, the high redshift limit in a given
redshift shell was used. The maximum detectable redshift was computed in 
the same fashion as in H92 (Eqn. 2 in their paper) using 
the sensitivity limits in Table 3 of H92 and either our new ROSAT 
determination of the cluster luminosity (using the same csomology and k--corrections) or that given in Table 2 of H92
(for clusters with no new ROSAT information). 

These individual cluster volumes were then converted into a XCLF 
by binning them as a function of luminosity according to

\begin{equation}
n(L) = \sum_{j=1}^{n} \frac{1}{V_{a,j}\, \Delta L},
\end{equation} 
   
\noindent where $n$ is the number of clusters in the luminosity bin
$\Delta L$ wide. Again, we use new ROSAT measurements 
of the clusters X--ray luminosities where appropriate. 
Before we discuss our new determinations of the EMSS XCLFs, we
note that we reproduced the binned H92 XCLFs as
shown in Figure 2 of their paper using only the data given in
Table 2 of H92. The re--determinations of the H92 XCLFs are shown in
Figure 7 for comparison and serve as a good check of our
methodology.

Meanwhile, Figure 8 shows our new determinations of the EMSS XCLFs in
the same three redshift shells as used by H92; z=0.14 to 0.2, 0.2 to
0.3 and 0.3 to 0.6.  We have used, where appropriate, the new data
presented in Table 1 of this paper, which includes removing clusters
that we have flagged as either unextended or uncertain (see the
discussion above and Table 1 where we have marked removed clusters
with a $\star$).  For clusters with no new information associated with
them, we simply used the original fluxes and luminosities quoted in
H92.  Poisson error bars are shown and were obtained from the
confidence limits tabulated by \cite{Gehrels86}.

\begin{table*}[thpb]
\begin{center}
\caption{
Maximum likelihood Parameterisation of XCLFs derived from
our New EMSS Data Sample}
\begin{tabular}[]{c|ccc} \hline
Redshift shell & Number & $\alpha$ & $K (10^7 Mpc^{-3}L_{44}^{\alpha-1})$  \\ \hline
  0.14 - 0.2 & 19 &$2.44\pm0.34$ ($2.19\pm0.21$) & $7.9\pm1.8$ ($5.85\pm0.25$)  \\
  0.2 - 0.3  & 21 &$2.66\pm0.30$ ($2.67\pm0.26$) & $6.5\pm1.7$ ($6.82\pm0.51$)  \\
  0.3 - 0.6  & 21 &$3.03\pm0.30$ ($3.27\pm0.29$) & $7.9\pm3.0$ ($12.33\pm3.87$) \\
  0.14 - 0.3 & 40 &$2.62\pm0.22$                 & $7.2\pm1.3$  \\ \hline
\multicolumn{4}{c}{With Luminosity Errors} \\ \hline
  0.14 - 0.2 & 19 &$2.60\pm0.37$  & $8.7\pm2.0$   \\
  0.2 - 0.3  & 21 &$2.73\pm0.31$  & $7.0\pm1.8$   \\
  0.3 - 0.6  & 21 &$3.08\pm0.32$  & $8.5\pm3.4$   \\

\end{tabular}
\end{center}
\end{table*}

Clearly, the three new XCLFs presented in Figure 8 are similar to 
those shown in H92 and Figure 7. This
indicates that the EMSS sample is internally robust, since
50\% of the clusters used in our new EMSS XCLFs have either a changed
classification (cluster or not), a new flux measurement (different by
up to 60\%), and/or a new redshift. Certain clusters have changed
substantially in luminosity, populating different bins in our XCLF
than they did in the H92 result.  This is certainly true in the
highest redshift shell where the 2 clusters MS0015.9+1609 and
MS0451.6-0305 have been elevated in total luminosity to occupy a
previously unoccupied bin centered at $2.5\times10^{45}\,{\rm
erg\,s^{-1}}$. We note here that other re--determinations of
the ROSAT flux for these two clusters are in good agreement 
with ours (Hughes, private communication, Neumann \& B\"ohinger 1996, Donahue \& Stocke 1995).

\begin{figure}[thp]
\centerline{\psfig{file=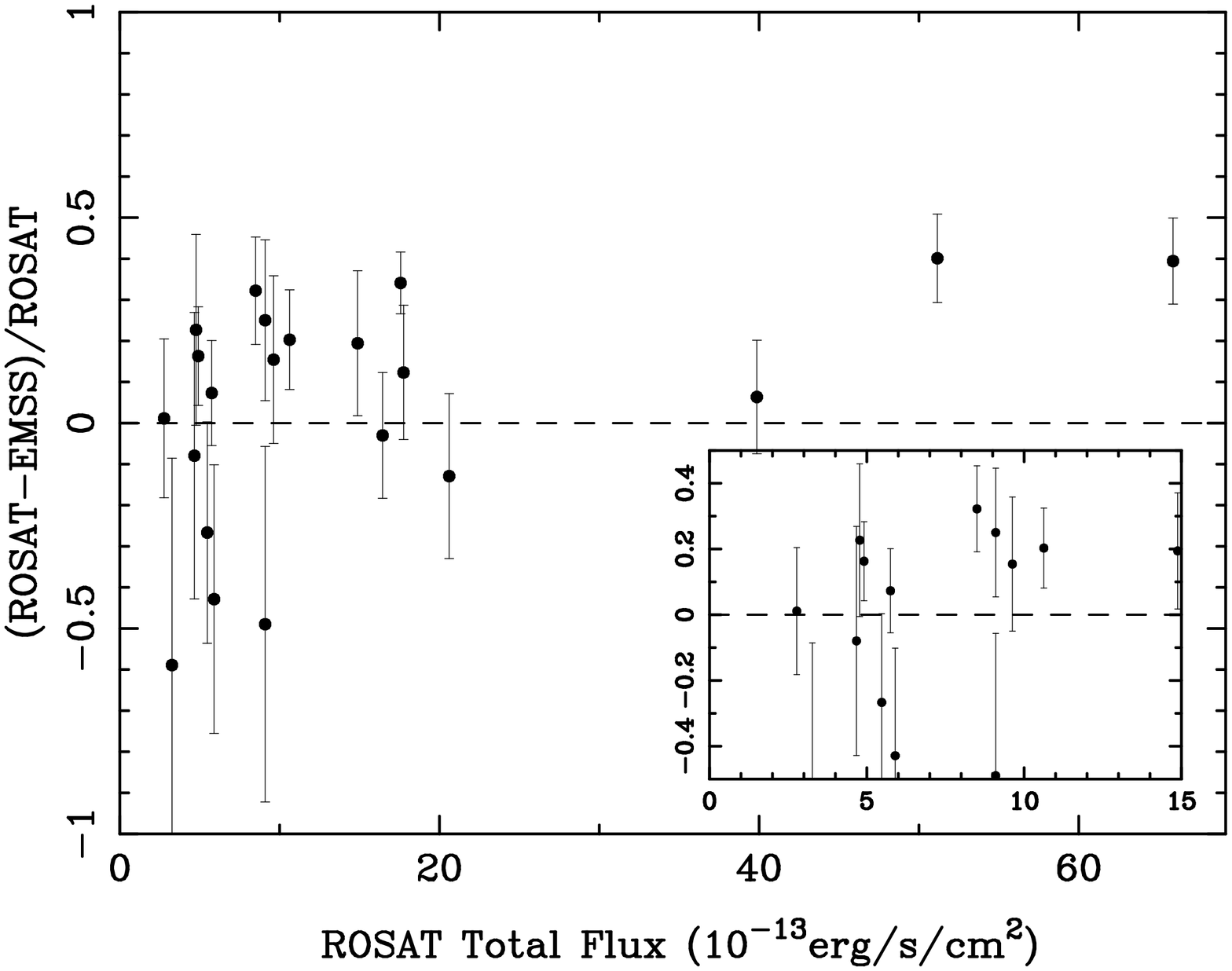,height=3.0in,angle=270}}
\caption{The percentile difference between our estimate of the EMSS
total flux (derived from our ROSAT data) and the total EMSS flux quoted by H92,
as a function of our estimate of the EMSS total flux. The errors plotted
were derived from the combinations of the errors on the net counts given 
in H92 and our errors in Table 1.} 
\end{figure}

\begin{figure}[thp]
\centerline{\psfig{file=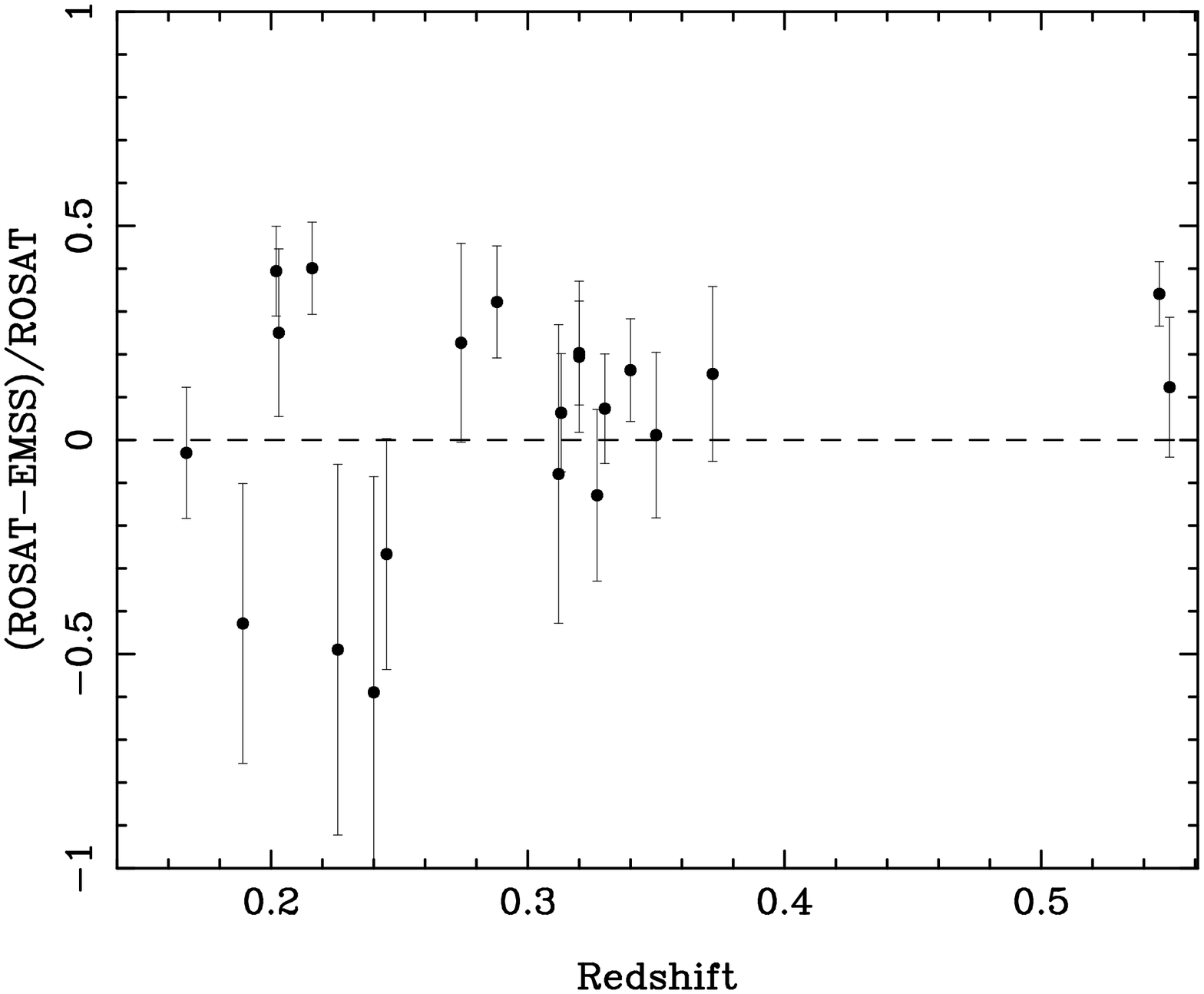,height=3.0in,angle=270.}}
\caption{The same as Figure 4, but plotted as a function of EMSS cluster redshift.} 
\end{figure}

However, these binned XCLFs do not allow us to judge the statistical 
significance of our work and therefore, we parameterised our luminosity
functions using a power-law luminosity function of the form,
$n(L_{44})=K\,L_{44}^{-\alpha}$, where $L_{44}$ is the clusters
luminosity in units of $10^{44}\,{\rm erg\,s^{-1}}$ (H92). For every cluster in a
particular redshift shell, we compute the probability that we would
see this cluster, at that luminosity, given the sensitivity limits of the
EMSS (in H92 Table 3) and the model. This probability ($P$) 
can be expressed in full as,

\begin{equation}
P(L_i) = \frac{1}{N} \int_{-\infty}^{\infty} V(L)\,L^{-\alpha}\,\delta(L_i-L)\,dL ,
\end{equation}

\noindent where $L_{i}$ is the observed luminosity 
of a cluster (taken from Table 1 of this paper or Table 2 from H92), 
$V(L)$ is the available volume within which that cluster could have been 
seen and $\delta(L_i-L)$ is a delta function
which is unity at $L_i$ and zero for all other values of $L$.
This, therefore, assumes no error on the observed luminosities (see below)
and Eqn. 3 can be reduced to 

\begin{equation}
P(L_i) = \frac{1}{N} V_a\,L_i^{-\alpha}\,\Delta L ,
\end{equation}

\noindent where $V_a$ is given in Eqn. 1 (the available search volume of the
cluster) and $\Delta L$ is chosen to be vanishingly small so only
a $L_i$ cluster can fall within this interval (see Cash 1979 for full
details).  In Eqn. 3 and 4, $N$ is the normalisation such that

\begin{equation}
N = \int_{L_{min}}^{\infty} \, V(L)\,L^{-\alpha}\, dL
\end{equation}

\noindent intergrated over the entire same range of luminosities. 
The likelihood of a given model is therefore,

\begin{equation}
{\cal L}(\alpha) = \prod_{i=1}^{n_{obj}} \, P(L_i|\alpha)
\end{equation}

\noindent
using Equations 3,4 and 5.

Since our ML method is slightly different from the approach
implemented by H92, we ensured that our ML fits to the original H92
data were compatible with those published in their paper.  In other
words, we attempted to fully reproduce the H92 result using only the
data in Table 2 of their paper.  We summarise these results in Table 2
and the fits are shown in Figure 7.  Overall, we obtained excellent
agreement with the H92 results with only one significant discrepancy;
we see a $50\%$ large error on the low redshift XCLF than H92 (In Tables 2 and 3, we present in parentheses the slopes, and errors, on all 3 EMSS XCLFs as 
published in H92).  The
source of this discrepancy remains unclear but may be the result of us
have a different likelihood distribution, for this low
redshift shell, than H92 and Gioia \etal (1990a). 
This would effect our error analysis because we integrated the
observed likelihood distribution directly, while H92 and Gioia \etal (1990a) 
used an analytical approach which assumes the likelihood distribution is a
Gaussian (Gioia, private communication).
We performed this latter approach
on our data and did find a smaller error on the slope of the low
redshift EMSS XCLF compared to that presented in Table 2. However, it was not
large enough to fully explain this discrepancy between us and H92.

\begin{figure}[thp]
\centerline{\psfig{file=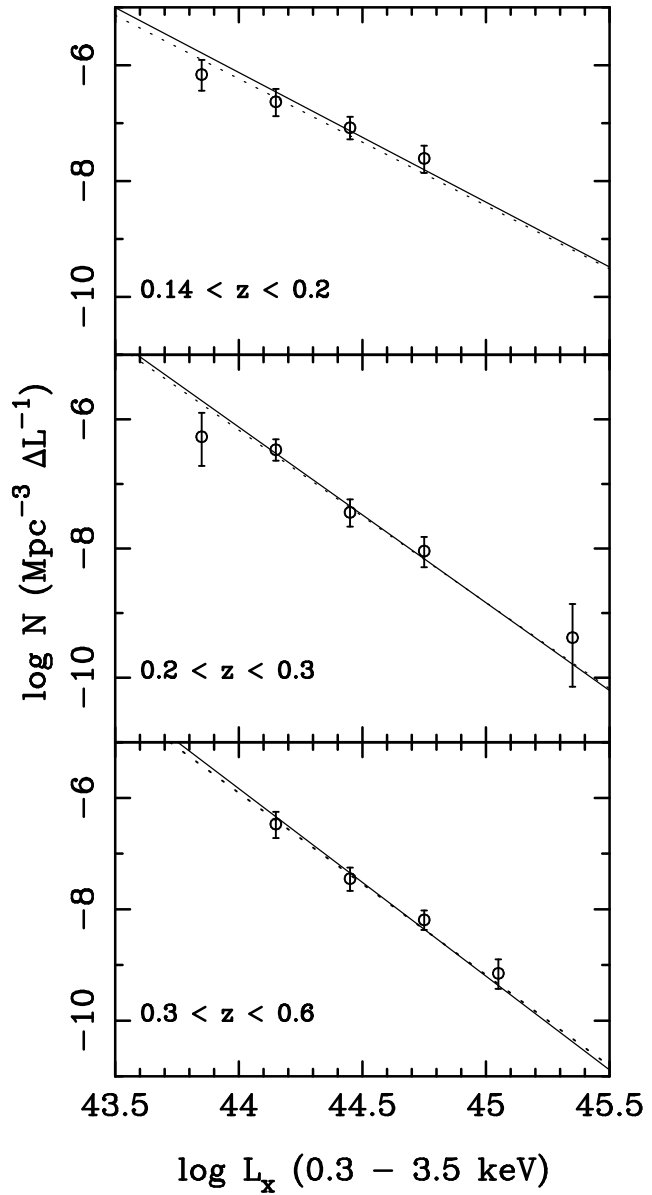,height=7.0in,angle=0.}}
\caption{Binned X--ray Cluster Luminosity Functions in three 
redshift shells. These were computed by us using only the original
data in H92 (we have used different bin centers to those used by H92).
The solid line represents our best ML fit while the dotted line is the
best ML fit from H92.}
\end{figure}

\begin{figure}[thp]
\centerline{\psfig{file=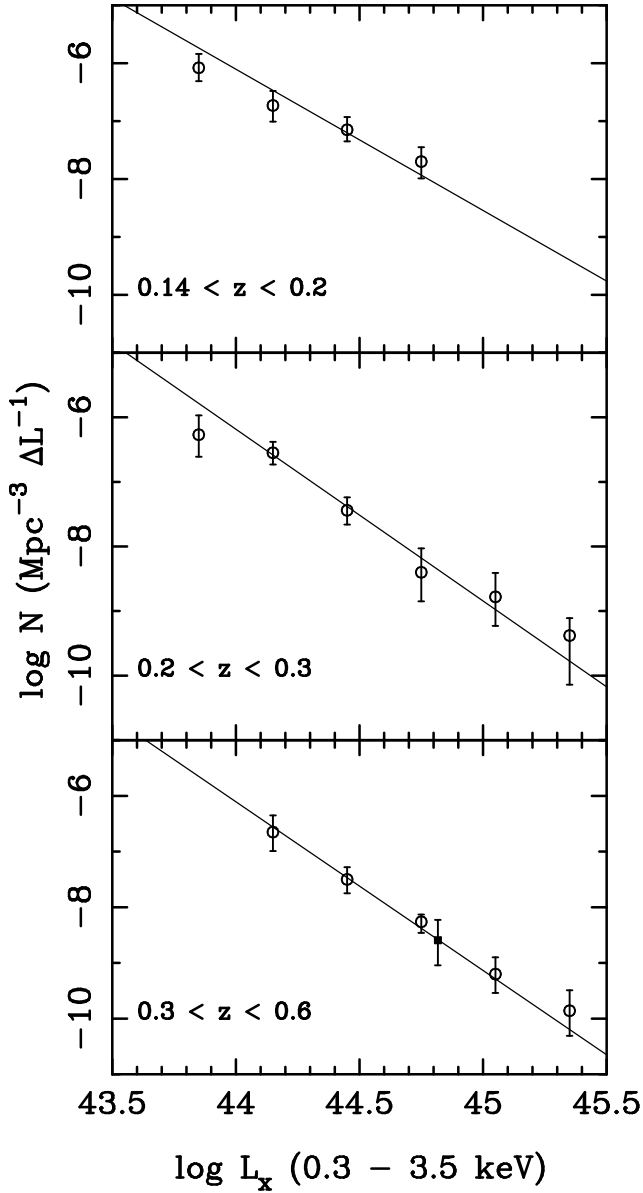,height=7.0in,angle=0.}}
\caption{Binned X--ray Cluster Luminosity Functions in three 
redshift shells. In this case, we have used the improvements detailed
in Table 1 of this paper in conjunction with data from the original
H92 data. The solid line represents our best ML fit as seen in Table
2.  The square symbol in the bottom plot represents the volume density
of very high redshift EMSS clusters discussed in the text.}
\end{figure}

Table 3 summaries the ML parameterisations of our new EMSS XCLFs and
includes (in parentheses) the original H92 results for comparison.  
The normalisation
of the power--law XCLF (K) was computed by requiring Eqn. 5 to give
the observed number of clusters.  Our ML parametric representations
are in reasonable agreement with those published by H92. This should
not be too surprising given the visual similarity in the binned XCLFs.
The largest discrepancy is in the slope of the highest redshift XCLF,
where we find a shallower slope than H92 by $\simeq 1\sigma$.  This
change is almost exclusively due to the increased dynamic range in
luminosity we are probing since we have a new highest luminosity bin
compared to H92.  We also find that our determination of the low
redshift XCLF has a steeper slope compared to the original H92 result
(we were able to re--produce the H92 slope when we used only their
data).  This is somewhat surprising since we have only changed 4
clusters in this redshift shell (2 changed in luminosity and 2
clusters were removed; see Table 1).  We have investigated this
discrepancy by re--computing the low redshift XCLF removing each of these 4
clusters one at a time. The most significant effect is noticed when
MS2318.7-2328 is removed because of a lower redshift measurement by
Romer (1994).  This cluster is the highest luminosity cluster 
in the original H92 determination of the low redshift XCLF and
therefore, it is understandable that its' removal has the largest 
effect on our XCLF re--determination (also it explains why we see a steeper slope
since we have removed the highest luminosity cluster in that shell).  
Overall, this low redshift shell is poorly determined since it
is very sensitive to small changes in the dataset.

To quantify the statistical significance of our ML fits, we present in
Tables 2 and 3 the computed errors on our ML parameters.  These were
computed by re--normalising the observed likelihood distributions (by
integrating over the whole distribution and dividing it by the total)
for the three redshift shells and determining the interval that
enclosed 68\% of the values.  Figure 9 presents our observed
likelihood distributions for the three redshift shells and can be 
approximated by a Gaussian. This figure is extremely
enlightening, since it shows that the slope ($\alpha$) of the low
redshift parametric XCLF is insufficiently constrained since it has a wide
likelihood distribution.  Overall, the
high redshift XCLF parameterisation is only mildly inconsistent with the 
low redshift data. The
probability of not getting the high redshift slope -- from the low
redshift data -- is only 93\%. This is caused by a combination of the
low dynamic range -- in luminosity -- that the low redshift data
extends over and the changes in the XCLF slopes mentioned above.
Moreover, the difference in slope between the low and middle redshift
XCLFs is clearly insignificant with the probability of obtaining the
middle redshift shell maximum likelihood slope -- from the low
redshift data -- being 51\%.

Since the error on the XCLFs is crucial to any measurement of XCLF 
evolution, we
tested our ML error analysis by performed Monte Carlo
simulations ($>2000$ iterations per XCLF). This involved generating
fake cluster databases -- with the same total number of clusters as
the real data -- drawn at random from the parameterised H92 XCLFs. For
all three redshift shells, the mean of the subsequent distribution of
fitted XCLF slopes measured for these fake cluster catalogues was
identical to the original input values. Moreover, the distributions
were Gaussian--like and had dispersions very similar to those
presented in Tables 2 and 3. Therefore, our error estimates 
on the real XCLFs given in Tables 2 and 3
are consistent with the scatter expected for these number
of clusters (in each redshift shell).

\begin{figure}[thp]
\centerline{\psfig{file=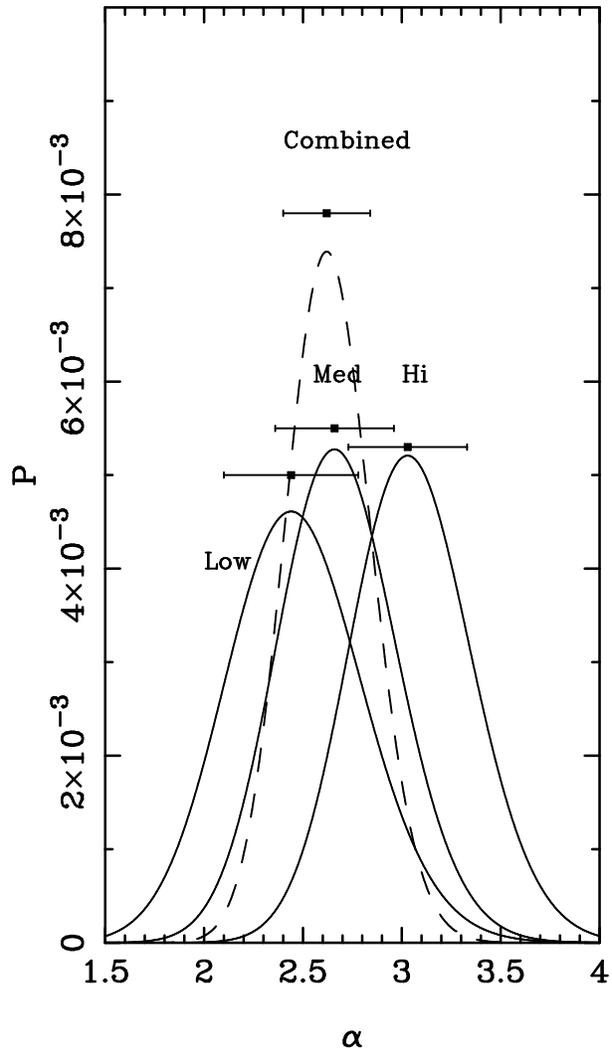,height=7.0in,angle=0.}}
\caption{The likelihood distributions 
for the three different redshift EMSS X--ray Cluster Luminosity
Functions. The dashed line is the combined luminosity function
($z=0.14$ to 0.3) discussed in the text. The solid points --with
errors-- represent the maximum likelihood slopes as tabulated in Table
2.}
\end{figure}

\begin{figure*}[thp]
\centerline{\psfig{file=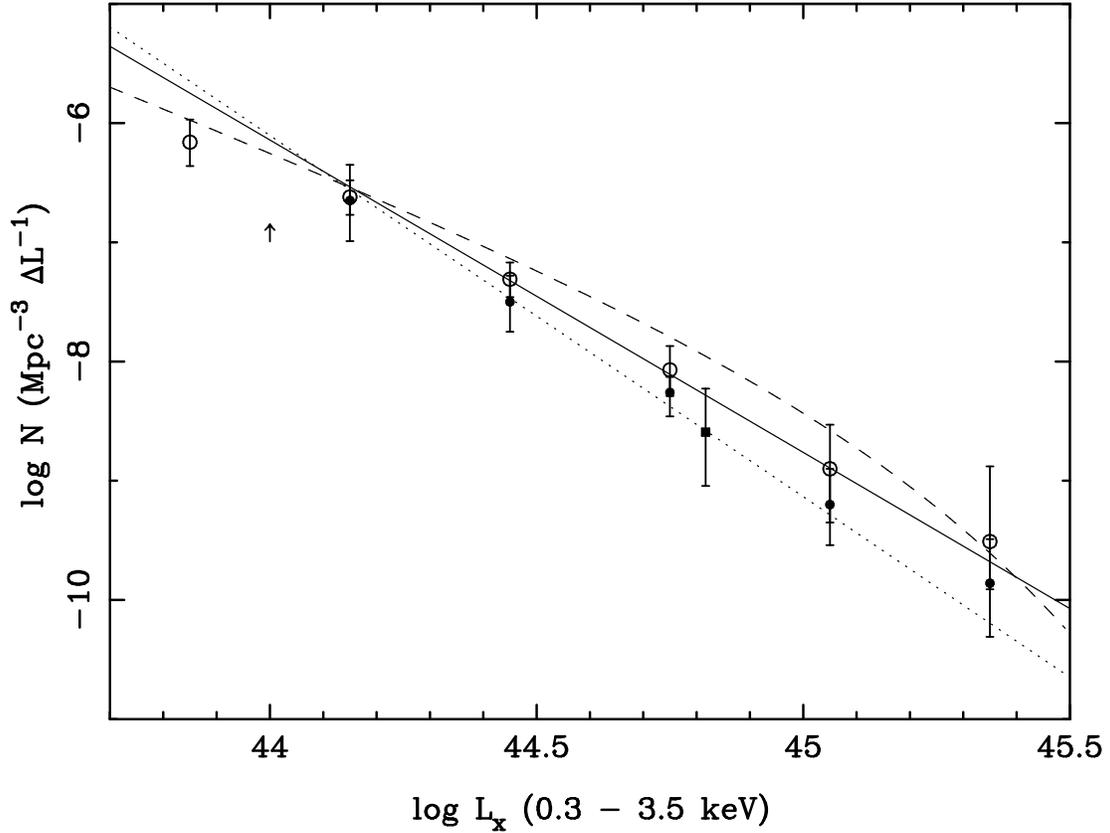,height=5.0in,angle=270.}}
\caption{Comparison of the low redshift XCLF (solid line) with
the high redshift XCLF (dotted line) and the XCLF of Ebeling \etal
(1995, 1996) computed from the RASS (dashed line). The latter has been shifted
to the same energy bandpass. The solid symbols ($\bullet$) are binned
high redshift luminosity function, while the open circles ($\circ$)
are the low redshift case. The solid square is the volume density of
very high redshift ($z>0.6$) clusters and the arrow ($\uparrow$) marks 
the lower limit of Nichol \etal (1994) in this redshift range.}
\end{figure*}

\begin{figure}[thp]
\centerline{\psfig{file=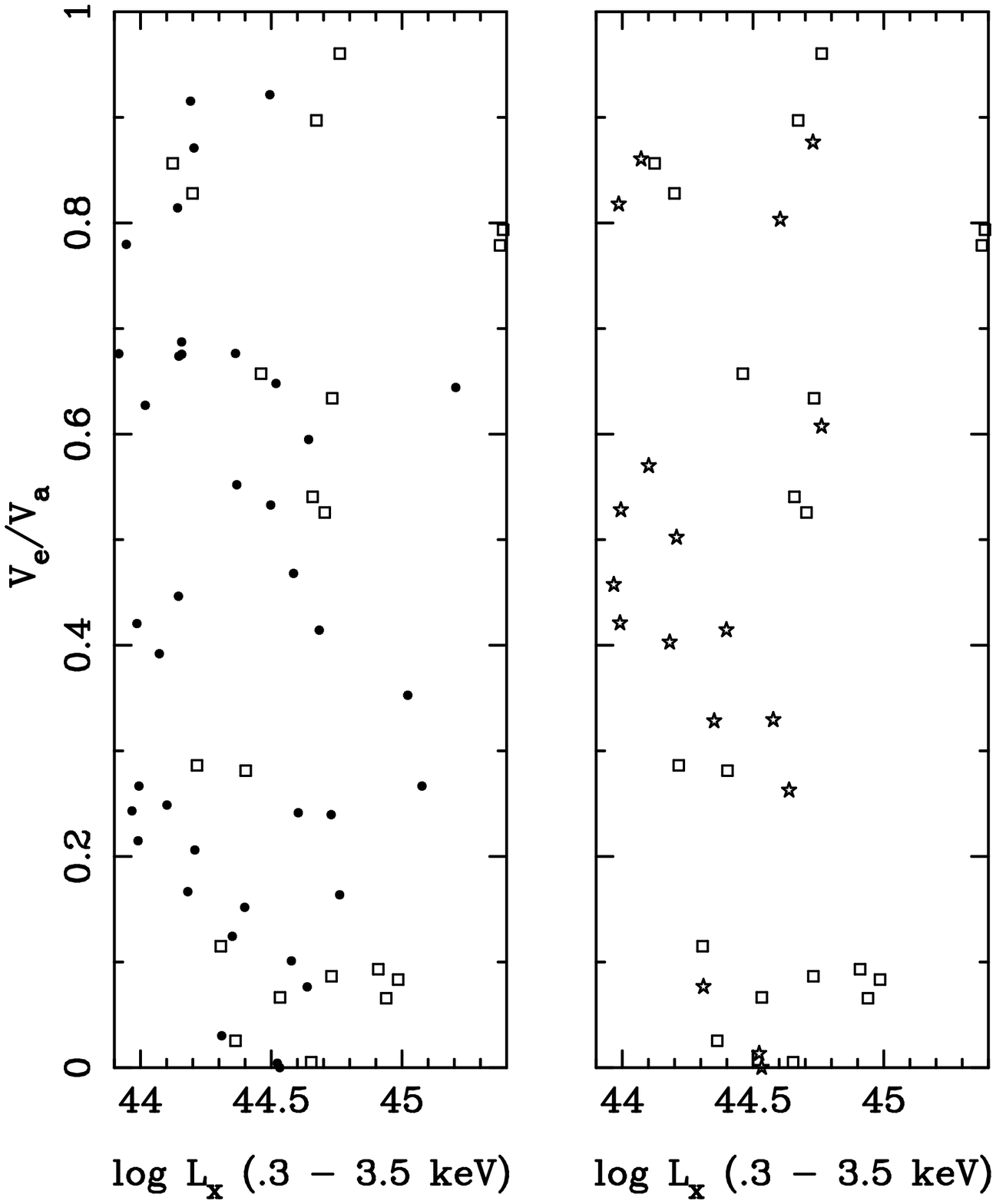,height=6.0in,angle=0.}}
\caption{$V/V_{max}$ versus $L_x$ for the EMSS sample. The left--hand
panel
is for the combined low (0.14 to 0.3) redshift shell ($\bullet$) against the high
redshift EMSS shell (open symbols). The right--hand panel is for the
original low
(0.14 to 0.2) redshift shell ($\star$) against the same high redshift
shell data.}
\end{figure}

\subsection{Further Tests of XCLF Evolution}

In this section we move beyond the analysis used by H92 to incorporate
different statistical tests and independent XCLF results published over the
last few years. We also investigated splitting the sample into
different redshift shells.

Our analysis has highlighted that the low redshift EMSS XCLF is insufficiently
constrained.  Recently, however, Ebeling \etal (1995, 1996) has published a new
low redshift XCLF using the ROSAT All--Sky Survey (RASS) Brightest Cluster
Sample comprising of 173 $z<0.3$ X--ray clusters.  From their
analysis, they found that the data was consistent 
with no XCLF evolution out to $z=0.3$.  This can be seen in
Figure 10, where we compare this new RASS XCLF with our re--determiations
of the EMSS XCLFs. We have now combined the lower two
redshift EMSS shells ({\it i.e.} $z=0.14$ to 0.3) since our analysis 
indicates that there is little statistical difference between the two.
Furthermore, it helps in our search for evidence for XCLF evolution
because it increases the dynamic range -- in cluster luminosities --
studied thus providing a greater constraint on the fitted ML
parameters (see Figure 9 and Table 3).

Clearly, there is excellent agreement between the RASS XCLF and our
combined lower redshift EMSS XCLF. This is very impressive since the two
samples cover different redshift regimes {\it i.e.} the RASS
sample only has $20\%$ ($8\%$) of its clusters at $z>0.14$ ($z>0.2$)  
-- the lower
limit of the EMSS data -- and all having $L_x>5\times 10^{44}\,{\rm
erg\,s^{-1}}$). The fact that these two XCLFs, from different surveys over different redshift ranges,
agree so well justifies our original motivation for combining the two
lower EMSS samples. Clearly, the emphasis should now be on determining
the degree of evolution above $z=0.3$. We say this irrespective of our
own work, since Ebeling \etal (1995, 1996) sees no evidence for evolution at
redshifts lower than this using a much larger sample of clusters. 
We have therefore, compared this new combined lower redshift EMSS
XCLF with that derived at high redshift.  Although the fitted ML
slopes of these two EMSS XCLFs are different ($z=0.14$ to 0.3 compared
to $z=0.3$ to 0.6), a close inspection of Figure 9 and Table 3 
highlights that this discrepancy is only mildly significant.

To quantify this significance, we have implemented a 2D Kolmogorov--Smirnov
(KS) test, in the $L_x$--$V/V_{max}$ plane, comparing the two redshift
shells, 0.14 to 0.3 and 0.3 to 0.6, from our improved EMSS sample. The
$V/V_{max}$ was calculated using the approach outlines in
Avni \& Bahcall (1980) using a uniform density distribution. This statistical analysis
removes the need for any binning, or fitting, of the data and directly
test the hypothesis that these two distributions were drawn from the
same underlying parent distribution.  Therefore, for this application, it is
a very powerful method.  Figure 11 shows the data displayed in the
$L_x$--$V/V_{max}$ plane for the two redshift shells mentioned above.
A KS test of these two gives a probability of 15\% that they were
drawn from the same underlying distribution.  This drops only slightly
to 11\% if we limit ourselves to comparing the $z=0.14$ to 0.2 shell
to the high redshift shell.  In these KS tests, we have constrained
the data to $L_x(0.3-3.5)>10^{44}\,{\rm erg\,s^{-1}}$ since below this
limit the high redshift shell starts to become seriously incomplete
(the EMSS does not have the sensitivity to probe such clusters at these
high redshifts). If we compare all the data without a luminosity cut, the 
above KS
probabilities -- that the two distributions are the same -- are
9\% and 5\% respectively.

For comparison, we also computed the mean $V/V_{max}$ in the high
redshift shell (over the same range of $L_{x}$) for both our improved
EMSS sample and the original H92 sample. We found $0.44\pm0.08$ for
our data and $0.36\pm0.06$ for H92 (the standard error on the mean is
quoted).  This again indicates that the evidence for evolution in the
high redshift shell has decreased since our measured mean $V/V_{max}$
is now consistent with the expected value (0.5; \cite{avni80}).

Most statistical analyses of XCLFs do not include 
observed luminosity errors. As can be seen
in Figures 5 and 6, this can be a large effect and a function of redshift. 
For completeness therefore, we 
re--computed our ML parameterisations of the XCLFs but replacing the delta
function in Eqn. 3 by a Gaussian whose width is equal to the observed 
luminosity error (using either the error on the flux given 
in Table 2 of H92 or our error on the ROSAT flux in Table 1 of this paper). 
This is computationally intensive since for each cluster the probability
of it being observed is now an integral over a range of possible luminosities
(because of the error in the luminosity). The search volume for each 
possible luminosity is re--computed (Eqn. 1) since $z_{max}$ changes with luminosity. 

In Table 3, we present these new ML parameterisation of the low, middle
and high redshift EMSS XCLFs. As can be seen, their are two
effects in adding such errors. First, all the slopes have steepened
with respect to their previous determinations (with no errors).
Second, the degree to which they have steepened is different for
each redshift shell, with the lower redshift shells changing the most.
Therefore, the statistical significance of any steepening with redshift
of these slopes is now lessened; the low and high redshift shells
are only inconsistent at the $1\sigma$ level.


Finally, we turn our attention to the form of the very high redshift
XCLF {\it i.e.} $z=0.6$ to $0.9$. In the past few years, several
groups have used the ROSAT satellite to extend the study of XCLF
evolution into this redshift regime (\cite{nichol94},
\cite{castander94}).  Moreover, further optical follow--up by GL94 has
revealed two new EMSS clusters that have $z>0.6$ (see also Luppino \& Gioia 1995). Although the data
are still very limited, we have plotted the volume density of these
two very high redshift EMSS clusters as a single point in Figure 10.
Also plotted is the lower limit for lower luminosity clusters in this
redshift shell as derived by \cite{nichol94}. In both case, these data
are consistent with the lower redshift EMSS XCLFs.

Using the EMSS sensitivity limits given by H92, we have computed that
the lowest detectable cluster luminosity in this very high redshift
shell is $4.1\times10^{44}\,{\rm ergs\,s^{-1}}$.  Given the low
redshift XCLF parameterisation (redshift 0.14 to 0.3; Table 3), we
estimate that the EMSS should have detected a total of 10 clusters
brighter than this limit at these very high redshifts (we would expect
three of these clusters to have $L_x>10^{45}$).  This discrepancy is
either evidence for evolution in the bright end of the XCLF at
these high redshifts, or, a reflection of the EMSS classification
procedure (the highest redshift clusters are the hardest to find and
measure).

\section{Conclusions}

We present in this paper a thorough re-examination of the EMSS cluster
sample using new data from the ROSAT PSPC pointing archive.
Furthermore, we include the latest information on the optical
follow--up of this sample of clusters. In total, 32 of the
original 67 $z>0.14$ EMSS clusters used by H92 to study X--ray cluster
evolution have new information associated with them, of which, 21 have
new X--ray data obtained from the ROSAT archive.  For these clusters,
we find no systematic difference, as a function of X--ray flux,
between the original EMSS flux estimates and our ROSAT
re--determinations ($28\%$ scatter). Our analysis did however,
indicate that this scatter may be correlated with cluster redshift
with the high redshift clusters having a lower scatter by a factor of
two. This is consistent with the expected trend given that the
original EMSS fluxes were computed within a fixed angular aperture.

We determined the extent for 21 of the ROSAT re-observed clusters, of
which, 14 were clearly extended compared to the ROSAT point--spread
function.  Combining this with results from H92, 40\% of the $z>0.14$
EMSS clusters are extended in the X--rays thus securing their
classifications as X--ray clusters. For the other 7 clusters, only 2
were definitely classed as point--like X--ray sources and the optical
follow--up work of these clusters supports the idea that these maybe
mis--classifications.  The remaining 5 clusters are uncertain, either
due to a lack of signal--to--noise to make a conclusive statement on
their extent, or, they simply resist easy categorisation (MS2137.3--2353,
MS1512.4+3647). 

This exercise also highlights the power of the SHARC analysis methods.  We
were able to extract known distant X--ray clusters from the 
ROSAT PSPC data archive based upon their observed X--ray extent
(clusters make up a small fraction of all X--ray emitting sources).
In cases where we did not detect an extent,  there is
a clear and well--understood reason for missing the cluster. 
For example, certain systems
($\sim 10\%$ of the EMSS) have intrinsically compact cores compared to
typical X--ray clusters and the rest of the EMSS sample and may
represent as a new class of
X--ray source (see Stocke \etal 1991).  Full details of the
SHARC survey selection
function will be presented in future papers.

We have used this improved EMSS catalogue to revisit the question of
X--ray cluster luminosity evolution. The original work of H92
suggested a steepening of the slope of the XCLF with redshift with a
statistical significance of $3\sigma$.  We employed similar
statistical analysis as H92 (binned $1/V_a$ test and Maximum
Likelihood analysis) to our data and found that, in general, we find
similar XCLFs as presented by H92.  However, we found two significant
differences.  First, Figure 9 shows that the EMSS provides a poor
determination of the local X--ray cluster luminosity function and 
shows that the low redshift EMSS XCLF is consistent with all
the other EMSS XCLFs.  Secondly, we find a shallower slope for the
high redshift EMSS XCLF, which immediately reduces the significance of
any claimed X--ray cluster luminosity evolution.

We have compared our determinations of the EMSS XCLFs with those found
in the literature.  For this comparison, we have combined the two
lower redshift EMSS XCLF since there is little statistical evidence
that these functions are significantly different. This combined low
redshift XCLF ($z=0.14$ to 0.3) agrees well with that derived from the
RASS (Ebeling \etal 1995, 1996) and strongly suggests there is no evidence for
XCLF evolution below $z=0.3$. Furthermore, a 2D KS test between this
combined low $z$ XCLF and the high $z$ XCLF indicates that the
significance of any evolution between the two is only 85\%. If we
compare the original low $z$ EMSS shell (0.14 to 0.2) with the high
$z$ shell, the significance of any observed evolution only rises to
$89\%$. These findings are now consistent with those of \cite{collins97}.

 At greater redshift ($z>0.6$), the data remains scarce, but at low
cluster luminosities the XCLF does not appear to be radically
different. At high luminosities ($L_x(0.3-3.5)>10^{45}\,{\rm
erg\,s^{-1}}$), the fact that the EMSS does not detect such clusters
may be indicative of evolution.  Overall, Figure 10 summaries our new
determinations of the EMSS XCLFs within the framework of other work in
this field and, we believe, demonstrates that the significance of any
claimed evolution with redshift can only be mild at best.  Several
projects are already underway ({\it i.e.} the SHARC survey;
\cite{burke95}) which will help clarify the situation by constructing
larger samples of X--ray clusters from the ROSAT data (both RASS and
archival data).

\section{Acknowledgements}

The authors would like to thank Francisco Castander, Harald Ebeling, Isabella
Gioia,
Jack Hughes, Avery Meiksin and Martin White for discussions, advice and help during
this work. We would also like to thank the combined efforts of Carlo Graziani,
Cole Miller and Jean Quashnock for their assistance and guidance on
the likelihood analysis presented in this paper. We thank Patricia Purdue for
extensive help with the X--ray reduction. We extend a special thank
you to the people at the GSFC ROSAT data archive and Steve Snowden for
his help and encouragement in adapting his data reduction software.
We are indebted to Michael Loewenstein for access to his propriety HRI data on MS1219.9+7542.
We thank an anonymous referee for their comments.  BH
acknowledges summer support at Northwestern University from a NASA
Space Consortium Grant through Aerospace Illinois. He also
acknowledges CARA for partial funding during this work. DJB and CAC
acknowledges PPARC for a studentship and Advanced fellowship
respectively. This work was partially supported by NASA ADP grant NAG5-2432.

\end{document}